\documentclass[doublecol]{epl2}
\usepackage{amsmath}
\usepackage{graphicx}% Include figure files
\usepackage{dcolumn}% Align table columns on decimal point
\usepackage{bm}% bold math
\usepackage{colordvi}
\usepackage{epsfig}
\usepackage{graphicx}
\usepackage{color}
\usepackage{booktabs}

\newcommand{\kap}{\boldsymbol{\kappa}}
\newcommand{\rb}{{\bf r}}

\newcommand{\sig}{\boldsymbol{\sigma}}

\title{Density Functional approach to Nonlinear Rheology}

\author{J. Reinhardt\inst{1} \and F. Weysser\inst{2} \and J.M. Brader\inst{1}}
\shortauthor{J. Reinhardt}
\institute{
\inst{1} Department of Physics, University of Fribourg, CH-1700 Fribourg, Switzerland\\
\inst{2} Institut Charles Sadron, CNRS, Strasbourg Cedex 2, France
}

\pacs{82.70.Dd}{Colloids}
\pacs{83.60.Rs}{Shear rate-dependent structure }
\pacs{05.20.Jj}{Statistical mechanics of classical fluids}
\abstract{
We present a density functional based closure of the pair Smoluchowski equation for Brownian particles 
under shear flow.
Given an equilibrium free energy functional as input the theory provides first-principles 
predictions for the flow-distorted pair correlation function and associated rheological quantities over a wide range of volume fractions and flow rates. Taking two-dimensional hard-disks under shear flow as an illustrative model we calculate the pair correlation function, viscosity and normal stress difference under both steady and start-up shear.
}

\begin{document}

\maketitle 

\section{Introduction}
The addition of colloidal particles to a Newtonian liquid gives rise to a
nonlinear rheological response, characterized by a rate dependent viscosity,
finite normal stress differences and nontrivial transient dynamics
\cite{larson1}. 
Understanding the interplay between particle interactions and external stress 
or strain fields remains a major theoretical challenge and much effort has been 
invested in the search for tractable closure relations which capture the essential 
physics of systems driven out-of-equilibrium \cite{brader_review}. Realistic 
models for which the competing effects of Brownian motion, potential and 
hydrodynamic interactions are simultaneously active pose particular 
difficulties \cite{dhont}.  

The microstructural distortion induced by an external flow field is encoded in
the nonequilibrium $n$-particle correlation functions.  For pairwise
interacting Brownian particles knowledge of the pair distribution function
alone is sufficient to calculate the full stress tensor, indicating that
specific features in this quantity can be correlated with nonlinearities in
the macroscopic rheology.  Such an approach was employed in recent experiments by
Cheng {\em et al.} \cite{cohen} and Koumakis {\em et al.} \cite{koumakis} in
which confocal microscopy was used to analyze the microstructural changes
associated with the onset of shear thinning, thickening and yielding.
Brownian and Stokesian dynamics simulations of hard spheres have provided further insight \cite{brady_review}, 
but a general theoretical method with firm thermodynamic foundation is still lacking. 
Existing theories focus either on states close to the glass transition by employing mode-coupling approximations \cite{brader1,brader2,fc09,brader3}, or attempt to extend exact low volume fraction results \cite{morris,bergenholtz} to finite volume
fraction using liquid state integral equation closures of the pair
Smoluchowski equation \cite{wagner,russel,lionberger,szamel,morris_superposition}.

In this paper we present a conceptually simple method by which the pair
correlations and nonlinear rheology of Brownian suspensions can be predicted
using dynamical density functional theory (DDFT) \cite{marconi,archer}. The accurate
treatment of packing effects and robust mean-field description of phase
equilibria provided by modern density functional approximations are inherited
by our theory, thus providing a simple and flexible framework within which the
interplay between particle interactions, diffusion and external driving may be
studied.  An advantage of our approach over existing integral equation based
theories is that we provide a clear link between the macroscopic rheology and
an underlying equilibrium free energy functional. The use of an explicit
generating function avoids the familiar problem of thermodynamic inconsistency
and no-solution regions of parameter space presented by integral equation
closures \cite{joe_ijtp}. This feature is of particular importance for systems
exhibiting equilibrium phase transitions, as the proximity of the chosen
thermodynamic state point to underlying phase boundaries may 
influence the rheological response. 
The present method is conceptually straightforward and may be applied to calculate the viscosity and normal stresses 
of any model for which there exists an explicit approximation of the free energy 
functional. 
It thus becomes possible to exploit the vast array of available equilibrium density functional 
approximations (for representative examples see \cite{AOfunctional,rod_needle,platelets,oettel}) 
for rheological studies.

\section{Theory}
We consider a system of $N$ Brownian particles interacting via isotropic
pairwise interactions, homogeneously dispersed in an
incompressible Newtonian fluid of given viscosity. In the absence of
hydrodynamic interactions the probability distribution of particle
positions is given by the Smoluchowski equation \cite{dhont}
\begin{eqnarray}\label{smol}
&&\hspace*{0.35cm}\frac{\partial \Psi(t)}{\partial t} + \sum_{i}\nabla_i \cdot {\bf j}_i =0,
\\
%\end{eqnarray}
%The probability flux of particle $i$ is given by
%\begin{eqnarray}
&&\hspace*{-0.5cm}{\bf j}_i = {\bf v}_i(t)\Psi(t) - D_0(\nabla_i - \beta{\bf F}_i)\Psi(t), 
\notag
\end{eqnarray}
where $D_0$ is the bare diffusion coefficient, $\beta\!=\!(k_BT)^{-1}$ and the
conservative force on particle $i$ is generated from the potential energy
according to ${\bf F}_i=-\nabla_i \sum_{j}u(r_{ij})$. To preserve
translational invariance we omit external potentials and consider a velocity
${\bf v}_i(t)=\kap(t)\cdot \rb_i$, where $\kap(t)$ is the traceless velocity gradient tensor.

Integration of (\ref{smol}) over the centre-of-mass of a pair of particles and
the remaining $N\!-\!2$ coordinates leads to the pair Smoluchowski equation
for the flow distorted pair distribution in which only the relative coordinate
$\rb_{12}=\rb_2\!-\!\rb_1$ appears
\begin{eqnarray}\label{pairsmol}
\hspace*{0.7cm}\frac{\partial g({\bf r}_{12})}{\partial t} + \nabla\cdot 
 {\bf j}(\rb_{12})
=0\,,
\end{eqnarray}
where $\nabla=\nabla_2=-\nabla_1$ and time arguments have been suppressed.
The pair flux consists of terms due to affine flow, Brownian motion, direct
and indirect interactions
\begin{eqnarray}\label{flux_exact}
\hspace*{0.1cm}{\bf j}(\rb_{12}) &&\hspace*{-0.3cm}= {\bf v}(\rb_{12})g(\rb_{12}) 
-\Gamma\Big[ k_BT\nabla g(\rb_{12})  + g({\bf r}_{12})\nabla u(r_{12}) 
\notag
\\
&& 
\hspace*{-0.9cm}
+\,\frac{\rho}{2} \!\!\int d\rb_3\, g^{(3)}(\rb_1,\rb_2,\rb_3)
\big( \nabla_2 u(r_{23}) - \nabla_1 u(r_{13}) \big)\Big],  
\end{eqnarray}
with $\Gamma\equiv 2\beta D_0$ and $\rho\equiv N/V$. The integral terms represent the residual influence of the particles which have been integrated out and require knowledge of $g^{(3)}(\rb_1,\rb_2,\rb_3)$.
%Note that the contribution $-\Gamma g({\bf r}_{12})\nabla u(r_{12})$,
%describing the direct interaction between the pair, formally can be regarded
%as a flux due to an external field $u(r_{12})$.
In the absence of flow the pair flux vanishes and (\ref{flux_exact}) reduces
to the exact second member of the YBG hierarchy \cite{hansen}.

\begin{figure}
\includegraphics[width=0.48\textwidth]{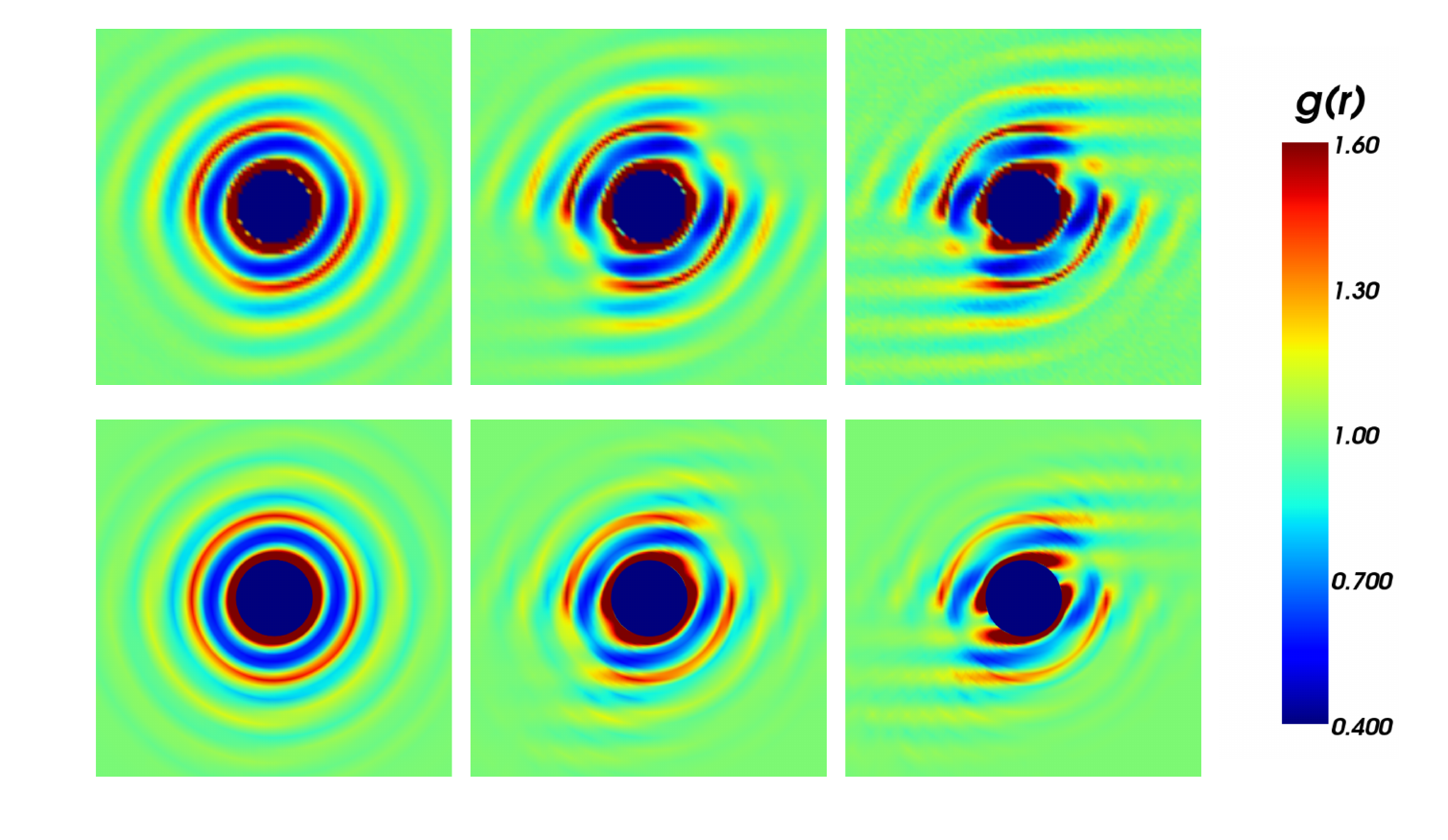}
\caption{
Comparison of pair correlation function $g(\rb)$ between simulation (top row) and DDFT (bottom row) at $\phi=0.6$. The Peclet number employed are $Pe=0.5$, $2$ and $5$, from left to right. 
The qualitative features are very well reproduced, although it is evident that the theory slightly 
underestimates the extent of the structural distortion.
}
\end{figure}	

The thermodynamic stress can be calculated directly from $g(\rb)$ by integration \cite{irving}
\begin{eqnarray}\label{stress} 
\sig(t)=-k_BT\rho \,{\bf 1} \,+\, \frac{1}{2}\,\rho^2\!\!\int \!d{\bf r} \frac{\rb\rb}{r}\, u'(r)g(\rb,t),
\end{eqnarray}
where we have made explicit the fact that all time-dependence comes from the
pair distribution.  It has been shown \cite{cohen} for low and intermediate rates that
supplementing the expression (\ref{stress}) by a volume fraction
dependent high frequency viscosity can account for the measured macroscopic
viscosity of hard sphere suspensions \footnote{
Although the non-hydrodynamic expression (\ref{stress}) was employed in \cite{cohen} 
to calculate the stress, hydrodynamic effects are nevertheless present in the 
experimentally measured $g(\rb)$ used as input.}.
%  For steady flows a
%detailed analysis of the low volume fraction limit, obtained by neglecting
%both the time derivative in (\ref{pairsmol}) and integral terms in
%(\ref{flux_exact}), has revealed that the stress integral (\ref{stress}) is
%dominated by a boundary layer in $g(\rb)$, within which Brownian and advective
%forces balance \cite{morris}. 

In order to close the theory we first note that the triplet distribution of
the considered homogeneous system may be related to an inhomogeneous pair
density 
\cite{shoulder}
\begin{eqnarray}\label{pair_density}
g^{(3)}(\rb_1,\rb_2,\rb_3) = \frac{\rho^{(2)}_{\,\rb_1}\!(\rb_2,\rb_3)}{\rho^2},
\end{eqnarray} 
where the subscript indicates that the source of inhomogeneity is the particle
located at $\rb_1$, now viewed formally as an external field perturbing the
density distribution.

An equilibrium sum rule may now be used to relate the r.h.s of \eqref{pair_density} to the gradient of the one-body direct correlation function \cite{archer}. 
\begin{eqnarray}\label{sum_rule}
-k_BT g(\rb_{12})\nabla_2 c_{\,\rb_1}^{(1)}(\rb_2)= 
\int \!d\rb_3\, \frac{\rho^{(2)}_{\,\rb_1}\!(\rb_2,\rb_3)}{\rho}\,\nabla_2 u(r_{23}),
\\
\notag
\end{eqnarray} 
where $c_{\,\rb_1}^{(1)}(\rb_2)$ is the direct correlation function at $\rb_2$
in an external field generated by the particle at $\rb_1$.  The assumption
that (\ref{sum_rule}) holds also in nonequilibrium constitutes an adiabatic
approximation \cite{reinhardt}, namely that the inhomogeneous pair density
relaxes instantaneously to that of an equilibrium system with density profile
$\rho g(\rb_{12},t)$.
Given the Helmholtz free energy
\begin{eqnarray}\label{helmholtz}
\mathcal{F}[\rho(\rb,t)]&&\hspace*{-0.3cm}=k_BT\int \!d\rb\, \rho(\rb,t)[\,\ln(\Lambda^3\rho(\rb,t))-1\,]
\notag
\\
&&\hspace*{-0.28cm}+\;\;\; \mathcal{F}_{\rm ex}[\rho(\rb,t)] \;+ \int \!d\rb\, u(r)\rho(\rb,t)\,,
\end{eqnarray}
with thermal wavelength $\Lambda$, the direct correlation function is
generated by differentiation of the excess contribution
\begin{eqnarray}\label{direct}
c^{(1)}(\rb)=-\beta\frac{\delta \mathcal{F}_{\rm ex}[\,\rho(\rb)\,]}{\delta\rho(\rb)}.
\end{eqnarray}
Employing (\ref{pair_density}-\ref{direct}) and exploiting the symmetry of the
direct correlation function enables us to approximate the integral term in
(\ref{flux_exact}), yielding a closed equation of motion
\begin{eqnarray}\label{ddft}
&&\hspace*{0.cm}\frac{\partial g({\rb,t})}{\partial t} 
\!=\!-\nabla\!\cdot {\bf j}(\rb,t)
\\
{\bf j}(\rb,t)
\!\!\!\!&=&\!\!\!{\bf v}({\bf r},t)g(\rb,t) - \Gamma g(\rb,t)
\nabla\frac{\delta \mathcal{F}[\rho(\rb,t)]}{\delta \rho(\rb,t)}
\label{fluxx}
\end{eqnarray}
Equations \eqref{helmholtz}, \eqref{ddft} and \eqref{fluxx} thus provide a closed theory for calculating the pair correlation function which requires only that the equilibrium excess free energy functional corresponding to the pair potential $u(r)$ is known. Given $g(\rb,t)$ the stress tensor, and thus the macroscopic rheology, 
can be obtained from \eqref{stress}.

In order to test our approach we consider a two dimensional system of hard
disks of radius $R$ subject to shear flow ${\bf v}(\rb,t)=\dot{\gamma}(t) y
\,\hat{\bf e}_x$.  
This can be considered as a rather demanding test case, as the discontinuous nature of 
the interaction potential leads to strong spatial variations in $g({\bf r})$ which require 
both an accurate equilibrium functional as input to the dynamical theory and precise 
numerical algorithms with sufficient resolution. 
We employ a recently proposed approximation for
$\mathcal{F}_{\rm ex}$ due to Roth {\em et al}.  \cite{oettel}, which ensures
that the $g_{\rm eq}(r)$ generated by our theory in the absence of flow is in
excellent agreement with simulation data.  
For steady flows the relative
importance of advection with respect to Brownian motion is quantified by the
Peclet number $Pe=\dot\gamma R^2/2D_0$, which emerges naturally from
(\ref{ddft}) upon scaling time with $2D_0$ and taking the disk radius as the
unit of length.
Working in two dimensions enables accurate numerical solution
of (\ref{ddft}) using only modest computational resources and yields for fluid
states a phenomenology very similar to that observed in three dimensions.

\section{Numerics}

The numerical treatment of the hard disk case poses several difficulties.
Firstly, we are primarily interested in the value of $g(\rb)$ at contact, i.e.
on a circle of radius $2R$ around the center of the fixed test particle (see
Eq.(\ref{stress})).  The requirement to have grid points located precisely on
discontinuities in $g({\bf r})$ is incompatible with uniform cartesian grids of
the type usually employed in DDFT calculations.  Implementation of a cartesian
grid would require the use of a very fine grid spacing to obtain results of
sufficient accuracy, resulting in high computational demands.  Secondly, the
no-flux boundary conditions imposed by the hard-disk test particle require
special treatment to ensure that the continuity equation (\ref{ddft}) is
respected; particle number should not drift significantly during the
time-evolution as a result of numerical errors.  Thirdly, the equation is
stiff when fine grids are employed, so care is required for the
time integration. 

To overcome the geometrical incompatibilities of spherical particles and
cartesian grids, we choose a finite-element-like discretisation, allowing great
flexibility, as the grid can be partially refined in regions of interest, such
as the region of rapid variation in $g(\rb,t)$ close to contact. 

What distinguishes the present problem from those occuring in standard
computational fluid dynamics is that the fundamental measure free energy
functional employed requires the evaluation of nonlocal spatial convolutions
which are challenging to evaluate when working on an unstructured grid: 
Fast-Fourier-Transform techniques cannot be used. 
An advantage of the FMT approach is that the weight functions are
density independent and have limited range.  It is thus possible to
precalculate the convolution, so that it can be evaluated at each iteration
with only a matrix-vector product and the solution of a sparse linear system,
operations that are highly optimized in modern numerical linear algebra
libraries. We have implemented this scheme using the finite element framework
deal.II \cite{deal.II}. 

\begin{figure}
\center{\includegraphics[width=0.4\textwidth]{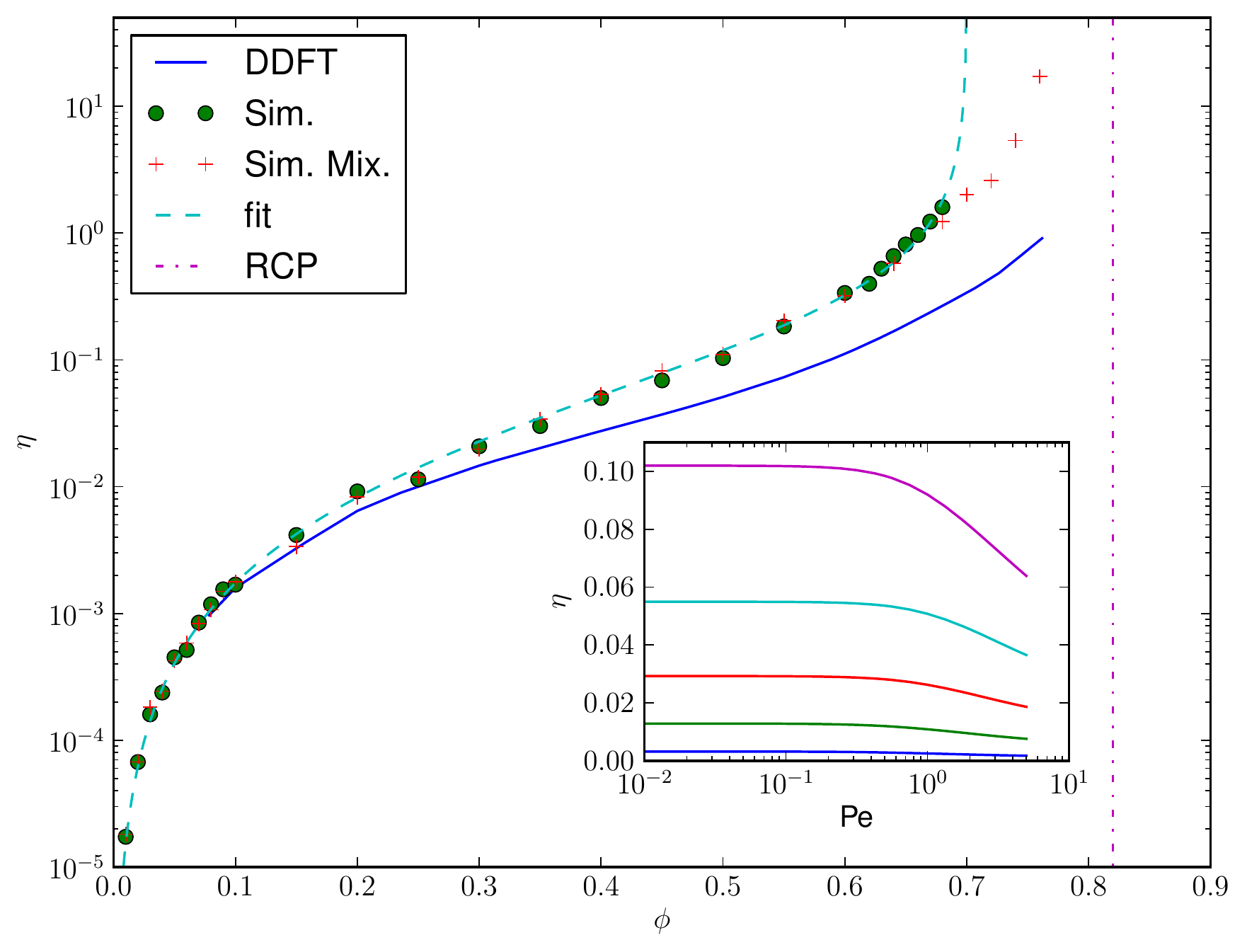}}
\caption{Zero-shear viscosity (full line) in units of $k_BT/2D_0$ as a function of volume fraction. 
Circles are simulation data for a monodisperse system and crosses are simulation data for a non-crystallizing 
binary mixture (see text).
The dot dashed line indicates an estimate for the random close packing of monodisperse disks
($\phi_{\rm rcp}=0.82$) \cite{berryman}. The fit to the monodisperse simulation data is given by $\eta = 0.109 \phi^2 (0.700 - \phi)^{-0.917}$.
Inset: The shear rate dependent viscosity as a function of Peclet number for volume fractions 
$\phi=0.1-0.5$, in steps of $0.1$.   
}
\end{figure}

To realise the hard walls both the density and the density flux have to
be discretized.  This enables accurate implementation of the boundary
condition that the flux normal to the hard boundary must vanish, but increases
the computational effort when compared to a discretization of the density
alone.
The resulting differential equation for $\rho(\rb,t)$ is stiff, i.e. its
Jacobian has eigenvalues with very large eigenvalues stemming from the
diffusion term \cite{hairerwanner}. This results in stability issues of the
numerical solution when using simple timestepping methods, unless very small
step sizes are chosen. The finer the grid is chosen, the more pronounced
the stiffness and the smaller the steps necessary for stable time integration,
dramatically increasing the computational demand. 
Often implicit Runge-Kutta Methods allow a efficient solution of stiff
differential equations, but in this case the large dimension of the problem and
its non-linearity prohibit the use of these methods. We thus make
use of a class of explicit Runge-Kutta Methods that are specifically tailored
for stiff equations \cite{anderson}.  
%This reduces the computational effort and
%enables the use of 
%larger time-steps.

\begin{figure}
\onefigure[width=0.48\textwidth]{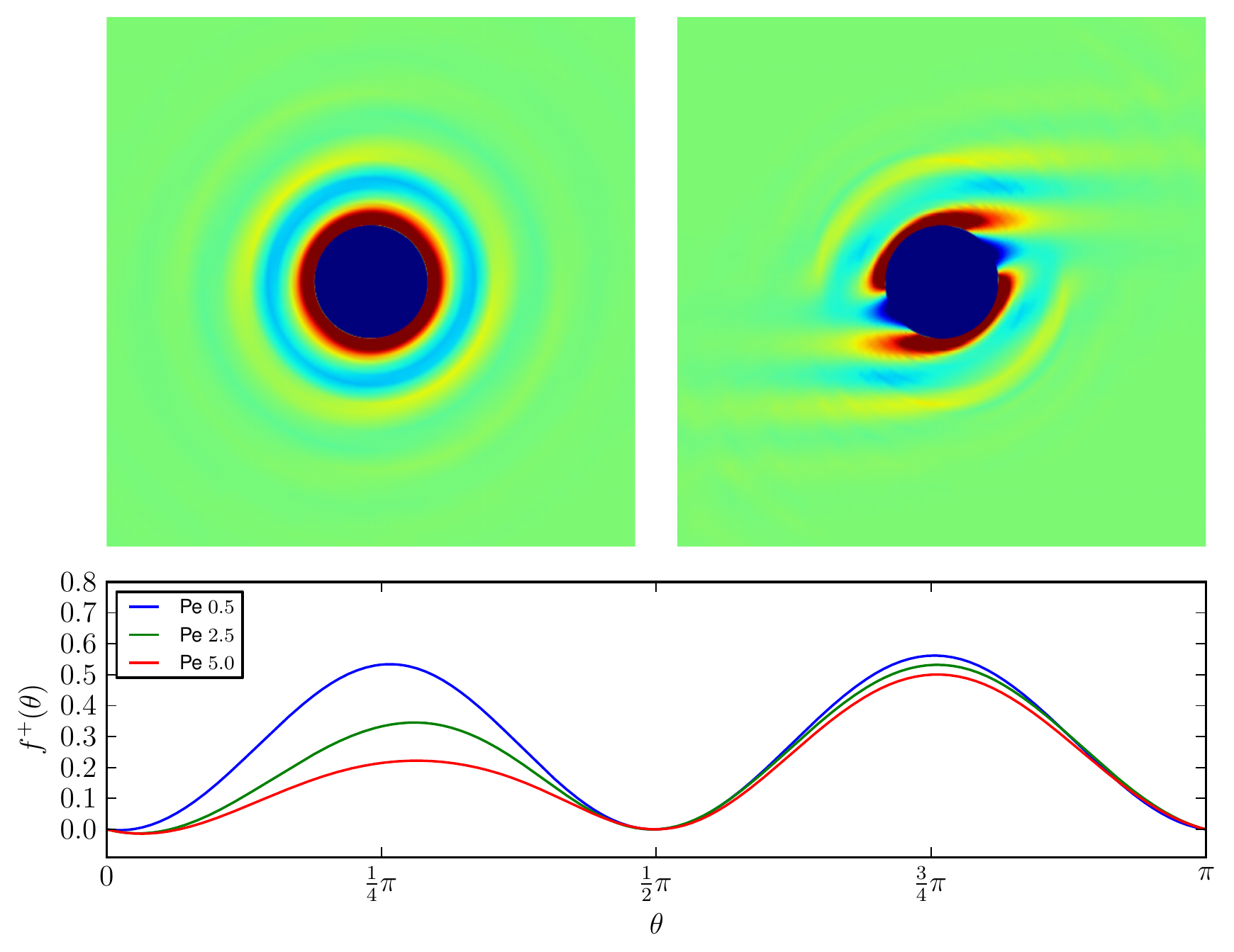}
\caption{
The pair distribution function for $Pe=0.5$ (top left) and $Pe=5$ (top right) at volume fraction 
$\phi=0.4$. 
We also show the angular dependence of 
$f^{+}_{\theta} \equiv \sin(\theta)\cos(\theta) (g_{eq}^+ - g_{\theta}^{+})/Pe$.
The viscosity for a given shear rate is given by the area under the corresponding curve. 
The angle $\theta$ is measured clockwise, starting from the top.
}
\end{figure}

\section{Simulations}

The simulation data were obtained using an event-driven algorithm whose detailed description for the non-sheared case can be found in \cite{weysser11}. Its extension to treat external fields is discussed in \cite{Henrich09}. We simulate a system of $N=1000$ hard disks with radius $R$ undergoing Brownian motion with the dimensionless diffusion coefficient $D_0/(2 v_0 R) =0.005$. Here $v_0$ is generated from a Gaussian velocity distribution imposed by the thermostat, 
ensuring that the distribution has the dimensionless variance of $m_0 v_0/(k_BT)=2$. The Brownian time as defined in \cite{weysser11} is set to $\tau_B v_0/d=0.01$.
Equilibration runs where performed, ensuring $\dot \gamma t \gg 1$ (cf.~\cite{Henrich09}) or $ D_0 t/R^2 > 10^3$(cf.~\cite{weysser11}) respectively.

In the simulation the pair distribution functions can be directly extracted from the particle positions. For the zero shear viscosity we use the collision based method from \cite{Henrich09} which avoids the cumbersome numerical evaluation of the limit $\dot \gamma \to 0$. The viscosity can hence be calculated via  
\begin{equation}
\eta_0 = \frac{1}{2 k_B T V} \lim \limits_{t \to \infty} \frac{1}{t} \left \langle \left( \sum \limits_{coll}  {\bf r}_y^{ij}  {\bf v}_x^{ij}  \right) \right \rangle,
\end{equation}
where $ {\bf r}_y^{ij}$ stands for the $y$-component of the relative coordinate between particle $i$ and $j$ and $ {\bf v}_x^{ij}$ for the $x$-component of the particles' relative velocity, while the sum runs over all collisions during the simulation time in the simulation box with the volume $V$. 

\section{Steady shear}

In Fig.1 we compare $g(\rb)$ from DDFT with Brownian dynamics data for a volume fraction $\phi = 0.6$ 
and three values of $Pe$. The chosen volume fraction lies well below the crystallization value of 
$\phi_{\rm cr} = 0.69$ but still represents a dense, strongly interacting colloidal liquid state.
At the lowest shear rate considered, $Pe=0.5$, the microstructure is only slightly distorted away from 
equilibrium and the ring structure of nearest, next-nearest etc. neighbour peaks is still clearly 
visible, reflecting the fact 
that Brownian forces are dominant over shear forces in this regime. 
As $Pe=0.5$ is a close-to-equilibrium state the good level of agreement between theoory and simulation 
owes much to the high quality of the excess free energy functional employed \cite{oettel}, which generates 
the equilibrium structure. The most noticable discrepancy is an overestimation of the higher order peak heights 
in the two extensional quadrants ($xy>0$).  
At the intermediate shear rate, $Pe=2$, shear forces start to dominate the Brownian forces and the microstructure 
becomes qualitatively different from equilibrium. 
Peaks in $g(\rb,t)$ which lie in the compressional quadrant increase in amplitude and are narrowed as particles 
begin to pile-up against the test particle. 
In contrast, the microstructure in the extensional quadrants becomes weaker due to particle depletion 
in the wake behind the test particle. 
Even at this relatively modest value of $Pe$ we observe the formation of lane-type structures in the extensional 
quadrants, arising from excluded volume packing constraints. 
At the highest shear rate considered, $Pe=5$, the microstructure is quite different from that in equilibrium 
with a very pronounced laning structure in the extensional quadrants. 
While the qualitative level of agreement between theory and simulation remains good, the packing structure appears 
to be underestimated by the theory, an effect which we attribute to the adiabatic approximation used 
to close equation (\ref{flux_exact}).

In Fig.2 we show the zero-shear limit of the viscosity $\eta_0 = \lim_{\dot{\gamma} \to 0} \sigma_{xy} \dot{\gamma}^{-1}$ as a function of volume fraction. For $\phi < 0.45$ the calculated $\eta_0$ is in good quantitative agreement with Brownian dynamics, but for larger values of $\phi$ the simulation data exhibits a rapid increase not captured by theory. This discrepancy is due to the onset of slow structural relaxation at high volume fractions which becomes partially lost upon making an adiabatic closure of the theory. 
For volume fractions in excess of $\phi_{cr} = 0.69$ crystallization becomes an issue.   
Although our theory is in principle sensitive to crystallisation (by virtue of the generating functional employed \cite{oettel}) 
the results presented here follow the disordered branch of the free energy, which in reality becomes metastable above $\phi_{cr}$. For large values of $\phi$ approaching random close packing ($\phi_{\rm rcp}\approx 0.82$ for disks 
\cite{berryman}) accurate numerical solution of \eqref{ddft} becomes difficult, but the available data suggests 
that $\eta_0$ diverges at a volume fraction considerably lower than unity, the value at which the input free 
energy functional becomes singular. For comparison we also include in Fig. 2 simulation data generated for a 
binary disc mixture in which half of the particles have diameter unity and the other half diameter $1.4$. 
This mixture is known not to crystallize \cite{Henrich09} and so the close agreement with the monodisperse 
simulations suggests that neglect of crystallization in our theory is not responsible for the apparent discrepancy.  
The inset to Fig.2 shows the rate dependent viscosity from DDFT for several values of the volume fraction within 
the range $0 < \phi < \phi_{cr}$. In each case the Newtonian plateau is followed by a regime of shear thinning characterized by a volume fraction dependent thinning exponent.

In order to investigate the features of $g(\rb)$ responsible for the shear
thinning behaviour we plot in Fig.3 $g(\rb)$ at a fixed volume fraction, $\phi=0.4$, for a Peclet
number in the linear regime ($Pe=0.5$, left panel in the figure) and another in
the shear thinning regime ($Pe=5$, right panel in the figure).  In determining
the rheology of hard particle systems only the angular dependence of the
contact value $g^+_\theta$ around the particle surface is significant (see equation
(\ref{stress})). However, closer inspection reveals that it is the quantity
$f^{+}_{\theta} \equiv \sin(\theta)\cos(\theta) (g_{eq}^+ - g_{\theta}^{+})/Pe$,
which contributes to the viscosity in \eqref{stress}. In the lower panel
of Fig.3 we show $f^{+}$ as a function of $\theta$ (where
the angle is measured in a clockwise direction, starting from the vertical).
The angular variation of $f^{+}_{\theta}$ clearly shows the reduction in the height of the depletion peak at $\theta = \pi/4$ which, when integrated, leads to shear thinning. The density aggregation in the compressional quadrant at $\theta = 3 \pi / 4$ has only a very minor effect on the rheology.

\begin{figure}
\onefigure[width=0.48\textwidth]{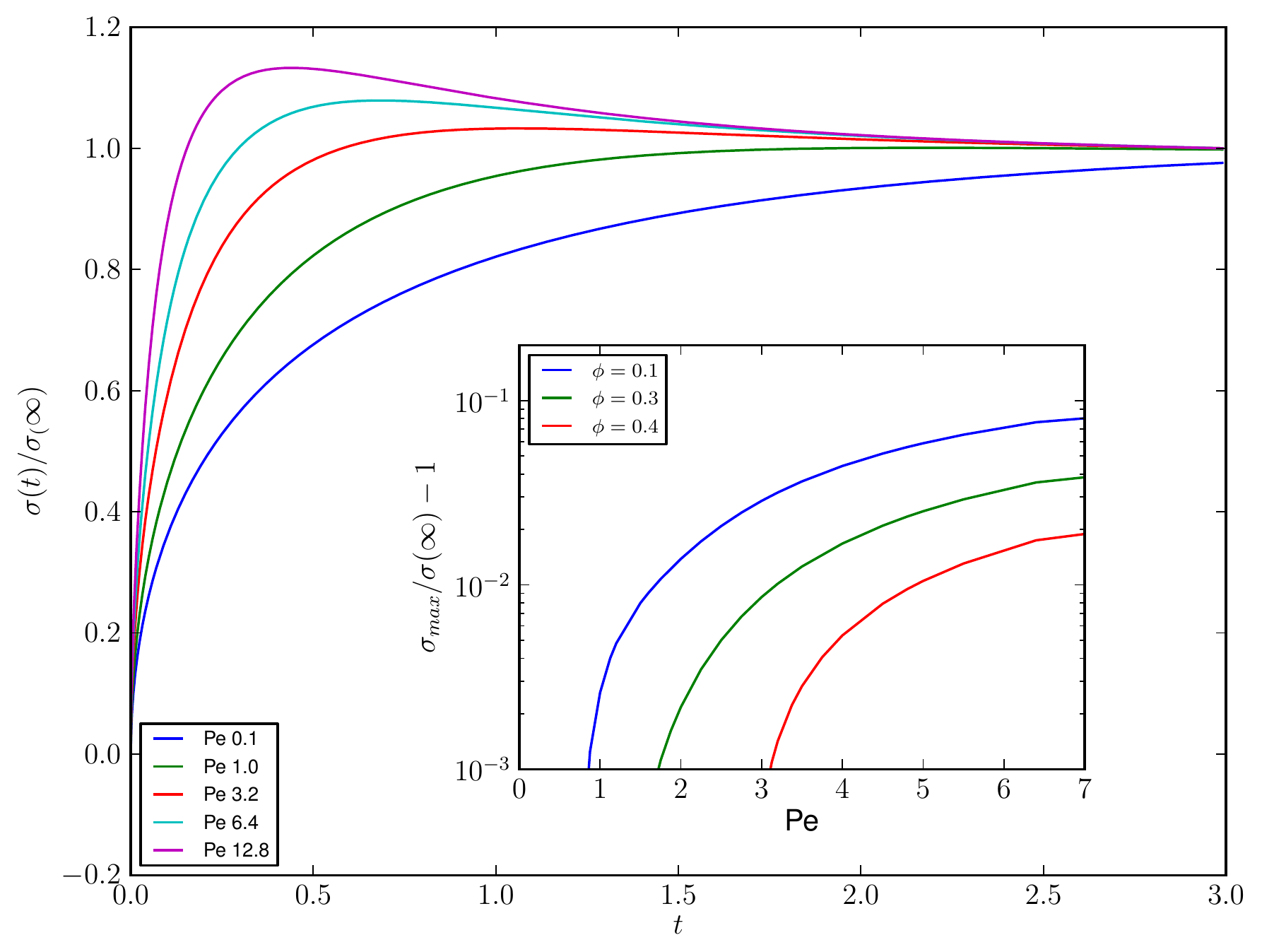}
\caption{
Stress curves for Start-up shear for $\phi=0.45$ and different $Pe$. Stresses are scaled by long time limit. For low shear rates the stress increases monotonically towards the long time limit. However for $Pe$ larger than a critical value there is an overshoot at short times whose relative magnitude increases with the shear rate. Inset: Relative magnitude of the overshoot for different packing fractions and shear rates. The overshoot increases with shear rate and decreases with volume fraction. 
%\comtwo{This is consistent with experimental findings by Koumakis et al}.
}
\end{figure}

\section{Startup shear}
In Fig. 4 we show the shear stress following the switch on of steady shear, $\dot{\gamma}(t) = \dot{\gamma} \Theta(t)$. This time-dependent shear field generates a transient response as the system evolves from equilibrium to the 
steady-state \cite{zausch}.
For small shear rates the stress monotonically approaches its long time value, but for more rapid 
shearing it first increases to a maximum value before slowly decaying to the steady state. 
The stress overshoot is a well known effect related to the passage from a regime of primarily elastic 
response to one of dissipative steady-state flow (see \cite{koumakis} and \cite{amann} for recent investigations of 
this phenomena). 
Both numerical simulations and experiment have found that the height of the overshoot maximum, relative to 
the steady-state shear stress, increases as a function of shear-rate but decreases with density \cite{koumakis,amann}. 
The former effect is a rather trivial consequence of appliying a stonger driving force, whereas the latter effect 
arises from the fact that suspensions at larger volume fraction have a smaller average distance between particle 
surfaces. Systems foor which the particles are densely packed begin plastic flow at smaller values of strain 
and are thus unable to accumulate large elastic stresses prior to the onset of flow.   
Our numerical results, shown in the inset to Fig. 4, are entirely consistent with this physical picture.
Also consistent with the findings of Koumakis {\em et al}. is the fact that the critical Peclet number for which 
the overshoot first occurs is density dependent (see inset Fig. 4). 

\begin{figure}
\onefigure[width=0.48\textwidth]{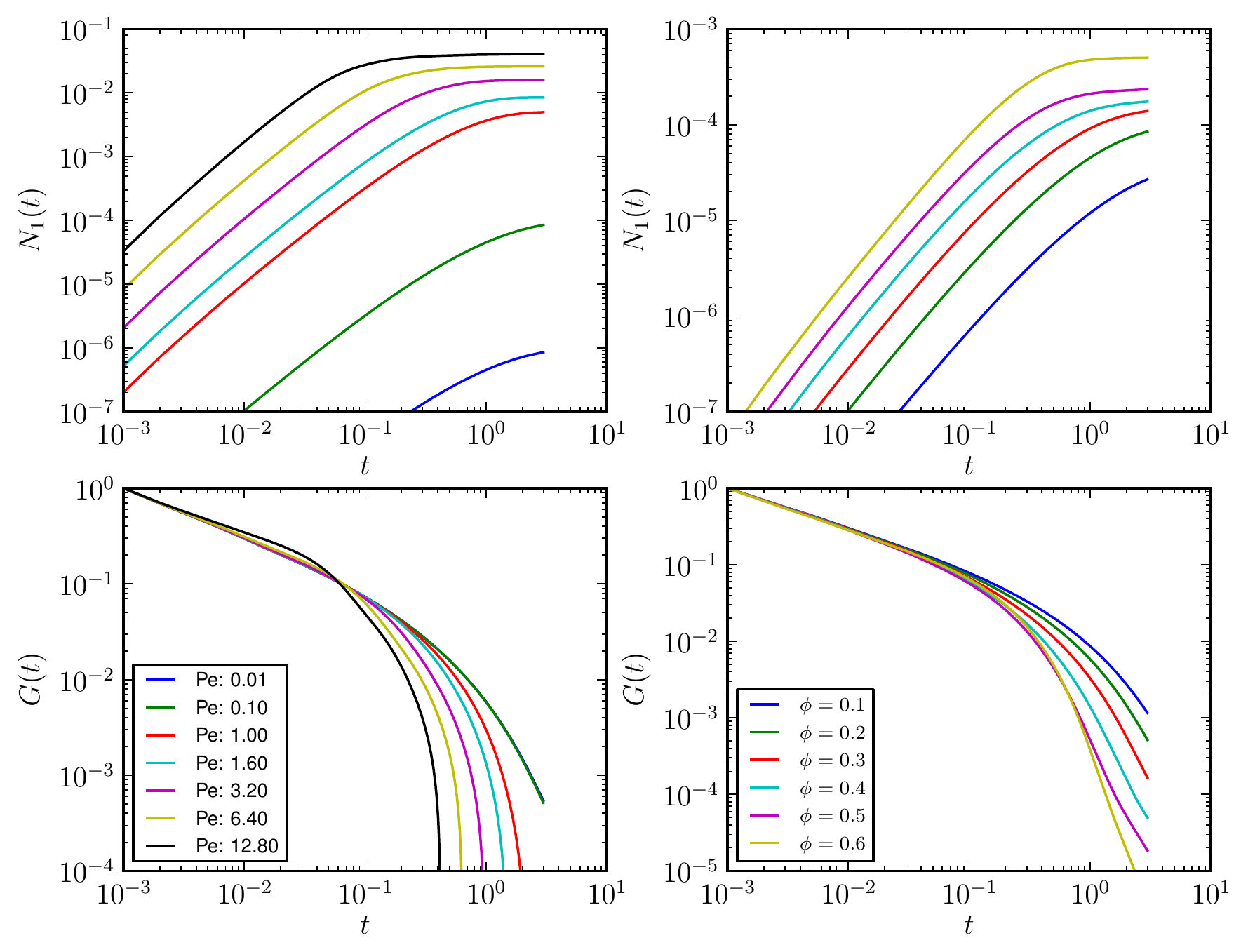}
\caption{
Buildup of normal stress differences (top row) and shear modulus $G(t)$ obtained from start-up (bottom row) for various shear-rates at $\phi=0.20$ (left column) and for various packing fractions at $Pe= 0.1$ (right column). 
}
\end{figure}

In the upper two panels of Fig. 5 we show the evolution of the first normal stress difference $N_1=\sigma_{xx}-\sigma_{yy}$ following 
start-up shear flow for various shear rates at fixed volume fraction and for various volume fractions at fixed 
shear-rate. The general trends follow closely those of the shear stress, but the overshoot is greatly suppressed 
and is not visible on the scale of the figure for the shear rates considered.  
The shear stress following switch-on shear also allows to obtain the shear modulus simply by differentiation 
with respect to time \cite{brader_review}. 
It can be seen that the relaxation time of the modulus decreases both with increasing shear and increasing packing fraction. 
%This is consistent with Green-Kubo relation for the viscosity. 
The bottom right panel shows the modulus for $\phi=0.2$. We note that at this volume fraction the 
shear modulus, $G(t)$, becomes negative at long times for $Pe> 1$ as a result of the stress overshoot in the stress.

\section{Summary}

In summary, we have developed a test-particle approach to calculating the shear-distorted pair correlation function of 
a colloidal suspension. Integration of the anisotropy generates the stress tensor, from which the nonlinear 
viscosity and normal stress differences may be calculated without further approximation. The second normal stress difference $N_2\equiv \sigma_{xx}-\sigma_{yy}$ was not considered explicitly here, due to our choice to focus on 
a two-dimensional model system, but is preducted by the theory when applied to three-dimensional systems. 
Furthermore, analysis of our fundamental equations, \eqref{helmholtz}, \eqref{ddft} and \eqref{fluxx}, shows that 
in three dimensional calculations there will be a nonvanishing distortion of $g(\rb)$ in the vorticity direction. 
Capturing this nontrivial distortion of $g(\rb)$ has been highlighted as an important challenge for theories of the 
nonequilibrium pair correlation functions \cite{szamel}.  
Due to our rather advanced algorithms for solving the DDFT equations we believe that full three 
dimensional calculations are entirely feasible and numerical studies of more realistic model fluids are the subject 
of further investigation.

The approach taken in the present work is close in spirit to the micro-rheological studies performed in \cite{krueger} 
and \cite{dzubiella}. While these works were focused on the density distribution about obstacles dragged through a 
suspension, we have focused on using the test particle method to obtain the bulk distorted pair-correlations as 
a gateway to macroscopic rheological studies. A key feature of our work is that the main rheological quantities 
are explicitly connected to an underlying free energy functional and that the method can be generally applied to 
investigate the rheology of any model system for which a free energy expression is available. 
The structural predictions of the theory for the difficult test case of hard discs are good when compared to 
the results of simulation (c.f. Fig. 1), however, discrepancies are clear in the zero-shear viscosity at intermediate 
and high volume fraction. 
For hard-discs this discrepancy can be attributed to the adiabatic closure approximation.  
By failing to recognize the existence of a glass-transition singularity at high density, the theory predicts too weak an increase of the viscosity as a function of volume fraction. 
It is interesting that even for monodisperse discs the glass transition has a dominant influence on the viscosity for 
volume fractions below freezing.  
Given that now there exist DDFT-type theories which promise to go 
beyond adiabaticity \cite{brader_krueger1,brader_krueger2,power} there would seem to be potential to similarly improve 
upon the theory presented here. Work along these lines is ongoing.

\acknowledgements

We thank Roland Roth for useful discussions and the Swiss National Science Foundation 
for financial support.

\end{document}